# Observation of charge density wave transition in TaSe$_3$ mesowires

J. Yang,[1,2] Y. Q. Wang,[1] R. R. Zhang,[1] L. Ma,[1] W. Liu,[1] Z. Qu,[1] L. Zhang,[1] S. L. Zhang,[1] W. Tong,[1] L. Pi,[1,2] W. K. Zhu,[1,a)] and C. J. Zhang[1,a)]

[1]*Anhui Province Key Laboratory of Condensed Matter Physics at Extreme Conditions, High Magnetic Field Laboratory, Chinese Academy of Sciences, Hefei 230031, China*

[2]*Hefei National Laboratory for Physical Sciences at Microscale, University of Science and Technology of China, Hefei 230026, China*

[3]*Institutes of Physical Science and Information Technology, Anhui University, Hefei 230601, China*

a)Authors to whom correspondence should be addressed: wkzhu@hmfl.ac.cn or zhangcj@hmfl.ac.cn

The quasi-one-dimensional (quasi-1D) TaSe$_3$ attracts considerable attention for its intriguing superconductivity and possible interplay with nontrivial topology and charge density wave (CDW) state. However, unlike the isostructural analogues, CDW has not been observed for TaSe$_3$ despite its quasi-1D character that is supposed to promote Peierls instabilities and CDW. Here we synthesize TaSe$_3$ mesowires (MWs) using a one-step approach. For the MW of ~300 nm thick, a distinct CDW transition occurs at 65 K in the resistivity measurement, which has not been reported before and is further evidenced by the Raman characterization and susceptibility measurement. For comparison, we have also prepared bulk single crystal TaSe$_3$. Although no anomaly appears in the resistivity and magnetoresistance measurements, the carrier type detected by Hall effect varies from n-type to p-type below 50 K, suggesting a reconstruction of Fermi surface that could be associated with CDW. The enhancement of CDW in the MWs is attributed to the reduced dimensionality. TaSe$_3$ is demonstrated to be a promising platform to study the correlation and competition of CDW and superconductivity in the quasi-1D systems.

Low-dimensional materials possess intriguing electrical, mechanical and optical properties and various types of potential applications, including field effect transistors,[1,2] flexible electronics,[2] gas sensors,[3] and optoelectronics.[4] Two-dimensional (2D) transition metal dichalcogenide (TMDC) thin flakes and quasi-one-dimensional (quasi-1D) trichalcogenide mesowires (MWs) (with formula $MX_3$, where M = Nb, Ta or Zr and X = S, Se or Te) are typical low-dimensional materials of this kind.[5,6] As a representative $MX_3$-type material, $TaSe_3$ has been well known for its elusive superconductivity in the past decades.[7-11] Recently, growing interests have been paid to $TaSe_3$ for its exotic electronic properties, such as high breakdown current[12,13] and low electronic noise.[14] Moreover, it has been theoretically predicted to be a strong topological insulator, in which the coexistence of topological phase and superconducting phase makes $TaSe_3$ a possible topological superconductor.[15]

Another notable property of trichalcogenide compounds is the so-called charge density wave (CDW) state that has been confirmed in $NbSe_3$,[16] $TaS_3$[17] and $NbS_3$.[18] $TaSe_3$ is an exception, for which the CDW transition has not been observed, even though the 1D metallic state should in principle become unstable and facilitate the stabilization of CDW. Also, unlike most quasi-1D materials, the rarely seen superconductivity in $TaSe_3$ implies possible competition between superconductivity and CDW that is usually present in high-temperature cuprates,[19,20] iron arsenide superconductors,[21] and some 2D TMDCs. The CDW states can be tuned to superconducting states by means of pressure or intercalation, for instance, in the intercalated $1T$-TiSe$_2$ and ZrTe$_3$,[22,23] and pressure-tuned TbTe$_3$.[24] In addition, CDW could still coexist with superconductivity, which makes the superconductivity unconventional.[25] Thus, the realization of CDW in quasi-1D TaSe$_3$ will be of special interest and importance for the current intensive studies on nontrivial topological states, superconductivity and CDW. Indeed, some efforts have been made to promote the electronic instabilities and CDW state in bulk TaSe$_3$, e.g., via chemical doping or applying uniaxial stress.[26-28] However, the resistance anomaly was not obvious and the proposed CDW transition temperature showed almost no change when increasing doping concentration. The direct observation of CDW transition in TaSe$_3$ is still lacking.

In this letter, we synthesize TaSe$_3$ MWs using a one-step approach. For the MW of ~300 nm thick, a distinct CDW transition occurs at 65 K in the electrical resistivity measurement, which is observed for the first time and further supported by the Raman characterization and magnetic susceptibility measurement. For comparison, we have also prepared bulk single crystal TaSe$_3$. Although no anomaly appears in the resistivity and magnetoresistance (MR) measurements, the carrier type detected by Hall effect varies from n-type to p-type below 50 K, suggesting a reconstruction of Fermi surface that could be associated with CDW. The enhancement of CDW in the MWs is attributed to the reduced dimensionality. TaSe$_3$ is demonstrated to be a promising platform to study the correlation and competition of CDW and superconductivity in the quasi-1D systems.

TaSe$_3$ MWs were prepared through a one-step method.[29] High purity (>99.99%, Alfa Aesar) tantalum and selenium powders in stoichiometry were mixed and ground. Then the mixture was sealed in an evacuated quartz ampule, heated to 600 °C at a rate of 2 °C/min, and maintained at this temperature for 20 h before cooling down to room temperature with furnace. Bulk single crystals of TaSe$_3$ were synthesized using a chemical vapor transport method. Stoichiometric power mixture were sealed in a quartz tube with iodine as transport agent (1 mg/cm$^3$). Ribbon-like single crystals were obtained after vapor transport growth for 10 days with a temperature gradient from 720 °C to 680 °C.

The crystal structure and phase purity were checked by X-ray diffraction on a Rigaku-TTR3 X-ray diffractometer using Cu Kα radiation at room temperature. The chemical component characterization was taken on an Oxford Swift 3000 energy dispersive spectrometer (EDS) equipped with a Hitachi TM3000 scanning electron microscope (SEM). Magnetic susceptibility was measured using a vibrating sample magnetometer (Quantum Design MPMS-3). Electrical measurements (electrical resistivity, Hall effect and *I-V* curve) were performed on a home-built Multi Measurement System (on a Jains-9T magnet). The SEM image of a typical device can be found in supplementary material. Raman spectroscopy was taken on a Horiba Jobin Yvon T64000 Micro-Raman instrument with a Torus laser (λ = 532 nm) as an excitation source in a backscattering geometry. The laser power was kept at 0.5 mW to avoid local heating effect.

The crystal structure of TaSe$_3$ is monoclinic with space group P2$_1$/m. In one unit cell, there are four linear chains along the *b* axis that can be divided into two groups, i.e., type I (orange) and type II (light blue), as shown in Fig. 1(a). In the chains, Se atoms (lavender) form trigonal prisms and Ta atoms (blue) are located at the prism center. Between the linear chains is the van-der-Waals gap, which facilitates the quasi-1D feature. Figures 1(c) and 1(d) show the distinctly different morphology of the bulk and MW of $TaSe_3$, respectively. The bulk single crystal is flat and ribbon-like, while the MW is slender, with a round cross section. The diameter of MW is ~300 nm and the length is ~8 cm (aspect ratio > $10^5$). Figure 1(b) shows the powder XRD pattern of the ground MWs, which indicates that the compound is single-phase $TaSe_3$ with lattice constants consistent with the literature.[30] We further check the structure of bulk $TaSe_3$ using single crystal XRD (supplementary material), which is also in good agreement with the space group $P2_1/m$. The chemical proportion of Ta : Se determined by the EDS is close to 1 : 3 (supplementary material). More SEM images of $TaSe_3$ MWs with various diameters ranging from 100 nm to 1 μm are included in supplementary material.

$TaSe_3$ has been reported to be a quasi-1D superconductor. However, the reported superconducting transition temperature are very different, ranging from 1.5 K to 2.3 K.[8,11,31] Meanwhile, the transition to the superconducting state depends strongly on the current density, which vanishes if the current density exceeds 3.0 A/mm$^2$.[31] Figure 2(a) shows the electrical resistance of bulk and MW samples (normalized by the resistance at 250 K), as a function of temperature from 250 K down to 2 K. The metallic behavior of bulk sample is similar to previous researches.[31] For the MW samples, we test a series of wires with different diameters to study the diameter dependence. The wire of 1.1 μm thick still exhibits a metallic behavior, like the bulk. For the 300 nm and 110 nm MW samples, an interesting kink (the resistance goes up with cooling) occurs in the curves. We define the onset temperature as $T_p$, which may be a phase transition temperature associated with CDW. The resistance increase is probably due to the gap opening of partial Fermi surface and the depression of density of states upon CDW transition. However, the rest ungapped Fermi surface can still induce conduction at low temperatures. Especially in the $MX_3$ quasi-1D materials, some chains can always be conducting without gapping. Therefore, the metallic behavior is restored with further cooling. Similar phenomena have been observed in the isostructural $MX_3$ materials, such as $ZrTe_3$,[23,32] $NbSe_3$,[29,33] etc. The $T_p$ is about 65 K and 150 K for the 300 nm and 110 nm MWs, respectively. The thinner wire has a higher transition temperature, suggesting that the CDW is enhanced by the reduced diameter. Further decreasing the diameter to 75 nm, the sample becomes more insulating and exhibits a completely semiconducting behavior. Hence for the 75 nm sample, the CDW cannot be determined by the resistivity measurement any more. From these measurements, the evolution of CDW with sample diameter becomes more clear. In the following text, we are focused on the 300 nm MW and bulk sample. Note that superconductivity is found in neither samples above 2 K, which may be due to the slight difference in stoichiometry of our samples from other superconducting samples.

Figure 2(b) shows the MR ($\frac{R-R_0}{R_0} \times 100\%$, where $R_0$ is the resistance at zero field) taken at 2 K under a magnetic field up to 8 T ($B \perp b$ axis). The MR is not saturated at high magnetic field, approaching 600% and 400% for the MW and bulk samples, respectively.

Such a giant positive MR is always related with nontrivial electronic topology,[34-39] which deserves more efforts in the future research.

In order to check possible field-induced effect in TaSe$_3$ MWs [nonlinear current-voltage (*I-V*) curves below $T_p$ is strong evidence for CDW], the *I-V* measurement was performed at 1.8 K [Fig. 2(c)]. The perfectly linear *I-V* curve indicates very good Ohm contact between metal electrodes and MW, which further confirms that the anomaly in *R-T* curve is intrinsic, ruling out the possible origin from poor contact. The inset shows the *I-V* curves of bulk sample taken between 2 K and 60 K, which are all linear. Figure 2(d) shows the *I-V* curve of MW taken at high electrical fields at 1.8 K. When the current reaches 5.55 mA, the dependence does not follow the Ohm's law but shows a cusp at the breakdown current density $J_B$=17.8 MA/cm$^2$, consistent with previous researches.[12] Such a large $J_B$ can be attributed to the unique quasi-1D chain structure which suppresses the electron scattering at grain boundaries and by interface dangling bonds.[40] The high current-carrying capacity suggests that TaSe$_3$ MWs have potential applications in microelectronics. Nevertheless, the nonlinear *I-V* curve is absent in our measurements. This may be because of the greatly enhanced depinning threshold field in nanoscale samples (usually several V or larger),[41,42] which obviously far exceeds the breakdown voltage of our sample [~0.3 V as seen in Fig. 2(d)].

There are several experimental techniques to identify CDW,[43] such as X-ray microdiffraction[44] and STM.[45,46] Raman spectroscopy is also a powerful tool to study the CDW in quasi-1D compounds and layered TMDCs through investigating the vibrational properties.[47-49] As seen in Fig. 3(a), all the peaks in the spectrum taken at room temperature are consistent with the literature,[12,50] again confirming the quality of our TaSe$_3$ MWs. Ten characteristic peaks can be found in the range of 100-300 cm$^{-1}$, originating from the primitive monoclinic structure of TaSe$_3$.[50] The peaks at 127 cm$^{-1}$ (B$_g$), 140 cm$^{-1}$ (A$_{1g}$), 164 cm$^{-1}$ (A$_{1g}$), 214 cm$^{-1}$ (A$_{1g}$), 238 cm$^{-1}$ (A$_{1g}$) and 260 cm$^{-1}$ (A$_{1g}$) are attributed to the out-of-plane modes, whereas the peaks at 109 cm$^{-1}$ (A$_g$), 120 cm$^{-1}$ (A$_g$), 176 cm$^{-1}$ (B$_2$/A$_g$) and 185 cm$^{-1}$ (B$_2$/A$_g$) are identified as the in-plane modes. Figure 3(b) shows the evolution of the Raman spectra of TaSe$_3$ MW with temperature. Interestingly, the peak at 120 cm$^{-1}$ (in-plane A$_g$ mode) shifts to high wave number with cooling, suggesting a possible phonon hardening process. Below 60 K, it merges with the 127 cm$^{-1}$ peak and remains unchanged until low temperatures. On the contrary, the out-of-plane B$_g$ (127 cm$^{-1}$) and A$_{1g}$ (238 cm$^{-1}$) modes show almost no changes with decreasing temperature. This is reasonable if we consider that the out-of-plane vibrations are not sensitive to the CDW transition as the in-plane vibrations. Furthermore, a new peak at 245 cm$^{-1}$ turns up at 60 K, corresponding to the transition to the CDW phase.

Figure 3(c) presents the temperature dependence of another in-plane mode, i.e., the 176 cm$^{-1}$ peak at 300 K, which shows a similar behavior with the 120 cm$^{-1}$ peak and can be associated with the CDW transition.[51,52] The kink point is 65 K, highly consistent with the $T_p$ revealed in the resistivity measurement. The data points above 65 K can be fitted by a linear formula,

$$\omega(T) = \omega_0 + \chi T$$

where $\omega_0$ is the vibration frequency at zero temperature and $\chi$ is the first order temperature coefficient. The $\chi$ obtained from the fit is -0.017 cm$^{-1}$ K$^{-1}$, which is close to that of TiS$_3$ nanosheet,[53] a confirmed CDW material. This value is much larger than that of 2D materials (e.g., $\chi_{A_{1g}}$=0.0032 cm$^{-1}$ K$^{-1}$ for monolayer WSe$_2$),[54,55] consistent with the enhanced 1D character and reduced interlayer coupling in TaSe$_3$ MWs.

Bearing in mind that the CDW transition is probably a thermodynamic process, we further perform magnetic susceptibility measurement for TaSe$_3$ MWs. As shown in Fig. 4(a), the diamagnetic susceptibility (~2.09×10$^{-4}$ emu/mol Oe) keeps nearly constant in the range of 100-300 K. With further cooling, the Pauli paramagnetic contribution from conduction electrons becomes dominant and gives rise to a sharp increase in susceptibility. It is worth noting that a slight difference between the ZFC (zero field cooling) and FC (field cooling) curves appears in the low temperature region, although the magnetic field dependence of magnetization shows perfect linearity within ±7 T [Fig. 4(b)], which rules out the possibility of magnetic hysteresis. If we plot the susceptibility as a function of inverse temperature [inset of Fig. 4(a)], the ZFC/FC bifurcation can be more clearly resolved. We can see that the bifurcation starts exactly from 65 K, which is consistent with $T_\text{p}$. Therefore, such an anomaly in susceptibility is another evidence for the CDW transition. The correlation between susceptibility and CDW can be ascribed to the reduced contribution to magnetic states from conduction electrons, due to the partial gapping of Fermi surface in the CDW phase.[22,56-58]

In order to depict the Fermi surface modulation effect, we further carry out detailed Hall effect measurements on the bulk sample. While no distinct anomaly is observed in the MR measurements near $T_p$ [Fig. 4(c)], the slope of Hall resistance undergoes an abrupt change across 50 K [Fig. 4(d)], indicating that the dominant carriers change from n-type to p-type. Similar behavior has also been found in NbSe$_3$, around its second CDW transition at 59 K,[59,60] The change of carrier type in bulk TaSe$_3$ suggests a reconstruction of Fermi surface which may be related to CDW, although no signature is observed in the resistivity measurement.

Before ending this section, we propose a possible scenario to explain why the CDW transition is only observed in the MW samples. In the bulk sample, despite the same 1D chains as the building bricks, the inter-chain interactions closely bind the chains and form 3D crystal, in which the quasi-1D character is strongly suppressed. In contrast, for the MW samples, the reduced dimensionality and enhanced quasi-1D character favor the Fermi surface nesting, which further promotes the Peierls instabilities and induces the CDW transition. In addition, the reduced dimensionality can enhance the electron-phonon interactions and then make for stronger CDW states. For instance, the CDW transition temperature of *2H*-NbSe$_2$ is increased when the sample thickness is decreased.[61] So the electron-phonon interactions may also play an important role in the occurrence of CDW in TaSe$_3$ MWs.

In summary, we synthesize TaSe$_3$ MWs using a one-step approach. For the MW of 300 nm thick, a distinct CDW transition occurs at $T_p$=65 K in the *R-T* curve, which is observed for the first time and further supported by the Raman characterization and susceptibility measurement. For comparison, we have also studied the bulk TaSe$_3$. Although no signature of CDW is observed in the resistivity and MR measurements, the carrier type varies from n-type to p-type below 50 K, suggesting a reconstruction of Fermi surface that could be associated with CDW. The enhancement of CDW in the MWs is attributed to the reduced dimensionality. TaSe$_3$ is demonstrated to be a promising platform to study the correlation and competition of CDW and superconductivity in the quasi-1D systems. Further research can be focused on the optimization engineering work, such as increasing the CDW transition temperature by gating technique.

See supplementary material for more SEM images of MWs, SEM image of device and characterization data.

This work was supported by the National Key R&D Program of China (Grant Nos. 2016YFA0300404, 2017YFA0403600, and 2017YFA0403502), and the National Natural Science Foundation of China (Grant Nos. U1532267, 11674327, 11874363, 51603207, 11574288 and U1732273).

[1]W. Liu, J. Kang, D. Sarkar, Y. Khatami, D. Jena, and K. Banerjee, Nano Lett. **13**, 1983 (2013).
[2]T. Roy, M. Tosun, J. S. Kang, A. B. Sachid, S. B. Desai, M. Hettick, C. C. Hu, and A. Javey, ACS Nano **8**, 6259 (2014).
[3]H. Li, Z. Yin, Q. He, H. Li, X. Huang, G. Lu, D. W. Fam, A. I. Tok, Q. Zhang, and H. Zhang, Small **8**, 63 (2012).
[4]F. Xia, H. Wang, D. Xiao, M. Dubey, and A. Ramasubramaniam, Nat. Photonics **8**, 899 (2014).

[5] J. O. Island, M. Buscema, M. Barawi, J. M. Clamagirand, J. R. Ares, C. Sánchez, I. J. Ferrer, G. A. Steele, H. S. J. van der Zant, and A. Castellanos-Gomez, Adv. Opt. Mater. **2**, 641 (2014).

[6] V. Y. Pokrovskii, S. G. Zybtsev, and I. G. Gorlova, Phys. Rev. Lett. **98**, 206404 (2007).

[7] Y. Tajima and K. Yamaya, J. Phys. Soc. Jpn. **53**, 3307 (1984).

[8] M. Yamamoto, J. Phys. Soc. Jpn. **45**, 431 (1978).

[9] M. Morita and K. Yamaya, Jpn. J. Appl. Phys. **26**, 975 (1987).

[10] S. Nagata, S. Ebisu, T. Aochi, Y. Kinoshita, S. Chikazawa, and K. Yamaya, J. Phys. Chem. Solids **52**, 761 (1991).

[11] Y. Tajima and K. Yamaya, J. Phys. Soc. Jpn. **53**, 495 (1984).

[12] M. A. Stolyarov, G. Liu, M. A. Bloodgood, E. Aytan, C. Jiang, R. Samnakay, T. T. Salguero, D. L. Nika, S. L. Rumyantsev, M. S. Shur, K. N. Bozhilov, and A. A. Balandin, Nanoscale **8**, 15774 (2016).

[13] T. A. Empante, A. Martinez, M. Wurch, Y. Zhu, A. K. Geremew, K. Yamaguchi, M. Isarraraz, S. Rumyantsev, E. J. Reed, A. A. Balandin, and L. Bartels, arXiv:1903.06227.

[14] G. Liu, S. Rumyantsev, M. A. Bloodgood, T. T. Salguero, M. Shur, and A. A. Balandin, Nano Lett. **17**, 377 (2017).

[15] S. Nie, L. Xing, R. Jin, W. Xie, Z. Wang, and F. B. Prinz, Phys. Rev. B **98**, 125143 (2018).

[16] R. V. Coleman, G. Eiserman, M. P. Everson, A. Johnson, and L. M. Falicov, Phys. Rev. Lett. **55**, 863 (1985).

[17] G. Grüner, Rev. Mod. Phys. **60**, 1129 (1988).

[18] Z. Z. Wang, P. Monceau, H. Salva, C. Roucau, L. Guemas, and A. Meerschaut, Phys. Rev. B **40**, 11589 (1989).

[19] E. H. da Silva Neto, P. Aynajian, A. Frano, R. Comin, E. Schierle, E. Weschke, A. Gyenis, J. Wen, J. Schneeloch, Z. Xu, S. Ono, G. Gu, M. Le Tacon, and A. Yazdani, Science **343**, 393 (2014).

[20] G. Campi, A. Bianconi, N. Poccia, G. Bianconi, L. Barba, G. Arrighetti, D. Innocenti, J. Karpinski, N. D. Zhigadlo, S. M. Kazakov, M. Burghammer, M. Zimmermann, M. Sprung, and A. Ricci, Nature **525**, 359 (2015).

[21] V. B. Zabolotnyy, D. S. Inosov, D. V. Evtushinsky, A. Koitzsch, A. A. Kordyuk, G. L. Sun, J. T. Park, D. Haug, V. Hinkov, A. V. Boris, C. T. Lin, M. Knupfer, A. N. Yaresko, B. Buchner, A. Varykhalov, R. Follath, and S. V. Borisenko, Nature **457**, 569 (2009).

[22] E. Morosan, H. W. Zandbergen, B. S. Dennis, J. W. G. Bos, Y. Onose, T. Klimczuk, A. P. Ramirez, N. P. Ong, and R. J. Cava, Nat. Phys. **2**, 544 (2006).

[23] X. Zhu, H. Lei, and C. Petrovic, Phys. Rev. Lett. **106**, 246404 (2011).

[24] J. J. Hamlin, D. A. Zocco, T. A. Sayles, M. B. Maple, J. H. Chu, and I. R. Fisher, Phys. Rev. Lett. **102**, 177002 (2009).

[25] Y. I. Joe, X. M. Chen, P. Ghaemi, K. D. Finkelstein, G. A. de la Peña, Y. Gan, J. C. T. Lee, S. Yuan, J. Geck, G. J. MacDougall, T. C. Chiang, S. L. Cooper, E. Fradkin, and P. Abbamonte, Nat. Phys. **10**, 421 (2014).

[26] A. Nomura, K. Yamaya, S. Takayanagi, K. Ichimura, T. Matsuura, and S. Tanda, EPL **119**, 17005 (2017).

[27] A. Nomura, K. Yamaya, S. Takayanagi, K. Ichimura, and S. Tanda, EPL **124**, 67001 (2019).

[28] T. M. Tritt, E. P. Stillwell, and M. J. Skove, Phys. Rev. B **34**, 6799 (1986).

[29]Y. S. Hor, Z. L. Xiao, U. Welp, Y. Ito, J. F. Mitchell, R. E. Cook, W. K. Kwok, and G. W. Crabtree, Nano Lett. **5**, 397 (2005).

[30]E. Bjerkelund and A. Kjekshus, Acta Chem. Scand. **19**, 701 (1965).

[31]S. Nagata, H. Kutsuzawa, S. Ebisu, H. Yamamura, and S. Taniguchi, J. Phys. Chem. Solids **50**, 703 (1989).

[32]X. Zhu, B. Lv, F. Wei, Y. Xue, B. Lorenz, L. Deng, Y. Sun, and C.-W. Chu, Phys. Rev. B **87**, 024508 (2013).

[33]T. M. Tritt, A. C. Ehrlich, D. J. Gillespie, and G. X. Tessema, Phys. Rev. B **43**, 7254 (1991).

[34]J. X. Gong, J. Yang, M. Ge, Y. J. Wang, D. D. Liang, L. Luo, X. Yan, W. L. Zhen, S. R. Weng, L. Pi, C. J. Zhang, and W. K. Zhu, Chin. Phys. Lett. **35**, 097101 (2018).

[35]D. D. Liang, Y. J. Wang, C. Y. Xi, W. L. Zhen, J. Yang, L. Pi, W. K. Zhu, and C. J. Zhang, APL Mater. **6**, 086105 (2018).

[36]Y. Wang, J. H. Yu, Y. Q. Wang, C. Y. Xi, L. S. Ling, S. L. Zhang, J. R. Wang, Y. M. Xiong, T. Han, H. Han, J. Yang, J. Gong, L. Luo, W. Tong, L. Zhang, Z. Qu, Y. Y. Han, W. K. Zhu, L. Pi, X. G. Wan, C. Zhang, and Y. Zhang, Phys. Rev. B **97**, 115133 (2018).

[37]Y. J. Wang, D. D. Liang, M. Ge, J. Yang, J. X. Gong, L. Luo, L. Pi, W. K. Zhu, C. J. Zhang, and Y. H. Zhang, J. Phys.: Condens. Matter **30**, 155701 (2018).

[38]D. D. Liang, Y. J. Wang, W. L. Zhen, J. Yang, S. R. Weng, X. Yan, Y. Y. Han, W. Tong, W. K. Zhu, L. Pi, and C. J. Zhang, AIP Adv. **9**, 055015 (2019).

[39]J. Yang, W. L. Zhen, D. D. Liang, Y. J. Wang, X. Yan, S. R. Weng, J. R. Wang, W. Tong, L. Pi, W. K. Zhu, and C. J. Zhang, Phys. Rev. Materials **3**, 014201 (2019).

[40]A. Geremew, M. A. Bloodgood, E. Aytan, B. W. K. Woo, S. R. Corber, G. Liu, K. Bozhilov, T. T. Salguero, S. Rumyantsev, M. P. Rao, and A. A. Balandin, IEEE Electron Device Lett. **39**, 735 (2018).

[41]J. McCarten, D. A. DiCarlo, M. P. Maher, T. L. Adelman, and R. E. Thorne, Phys. Rev. B **46**, 4456 (1992).

[42]T. L. Adelman, S. V. Zaitsev-Zotov, and R. E. Thorne, Phys. Rev. Lett. **74**, 5264 (1995).

[43]X. Zhu, J. Guo, J. Zhang, and E. W. Plummer, Adv. Phys. X **2**, 622 (2017).

[44]A. F. Isakovic, P. G. Evans, J. Kmetko, K. Cicak, Z. Cai, B. Lai, and R. E. Thorne, Phys. Rev. Lett. **96**, 046401 (2006).

[45]C. Brun, J. C. Girard, Z. Z. Wang, J. Marcus, J. Dumas, and C. Schlenker, Phys. Rev. B **72**, 235119 (2005).

[46]M. P. Nikiforov, A. F. Isakovic, and D. A. Bonnell, Phys. Rev. B **76**, 033104 (2007).

[47]P. Goli, J. Khan, D. Wickramaratne, R. K. Lake, and A. A. Balandin, Nano Lett. **12**, 5941 (2012).

[48]S. L. Gleason, Y. Gim, T. Byrum, A. Kogar, P. Abbamonte, E. Fradkin, G. J. MacDougall, D. J. Van Harlingen, X. Zhu, C. Petrovic, and S. L. Cooper, Phys. Rev. B **91**, 155124 (2015).

[49]H. Wang, Y. Chen, M. Duchamp, Q. Zeng, X. Wang, S. H. Tsang, H. Li, L. Jing, T. Yu, E. H. T. Teo, and Z. Liu, Adv. Mater. **30**, 1704382 (2018).

[50]T. J. Wieting, A. Grisel, and F. Levy, Mol. Cryst. Liq. Cryst. **81**, 117 (2011).

[51]D. L. Duong, G. Ryu, A. Hoyer, C. Lin, M. Burghard, and K. Kern, ACS Nano **11**, 1034 (2017).

[52]J. C. Tsang, C. Hermann, and M. W. Shafer, Phys. Rev. Lett. **40**, 1528 (1978).

[53]A. S. Pawbake, J. O. Island, E. Flores, J. R. Ares, C. Sanchez, I. J. Ferrer, S. R. Jadkar, H. S. van der Zant, A. Castellanos-Gomez, and D. J. Late, ACS Appl. Mater. Interfaces **7**, 24185 (2015).

[54]D. J. Late, S. N. Shirodkar, U. V. Waghmare, V. P. Dravid, and C. N. Rao, ChemPhysChem **15**, 1592 (2014).

[55]D. J. Late, ACS Appl. Mater. Interfaces **7**, 5857 (2015).

[56]J. A. Wilson, F. J. Di Salvo, and S. Mahajan, Phys. Rev. Lett. **32**, 882 (1974).

[57]X. Sun, T. Yao, Z. Hu, Y. Guo, Q. Liu, S. Wei, and C. Wu, Phys. Chem. Chem. Phys. **17**, 13333 (2015).

[58]M. Mulazzi, A. Chainani, N. Katayama, R. Eguchi, M. Matsunami, H. Ohashi, Y. Senba, M. Nohara, M. Uchida, H. Takagi, and S. Shin, Phys. Rev. B **82**, 075130 (2010).

[59]R. M. Fleming, J. A. Polo, and R. V. Coleman, Phys. Rev. B **17**, 1634 (1978).

[60]N. P. Ong and P. Monceau, Solid State Commun. **26**, 487 (1978).

[61]X. Xi, L. Zhao, Z. Wang, H. Berger, L. Forró, J. Shan, and K. F. Mak, Nat. Nanotechnol. **10**, 765 (2015).

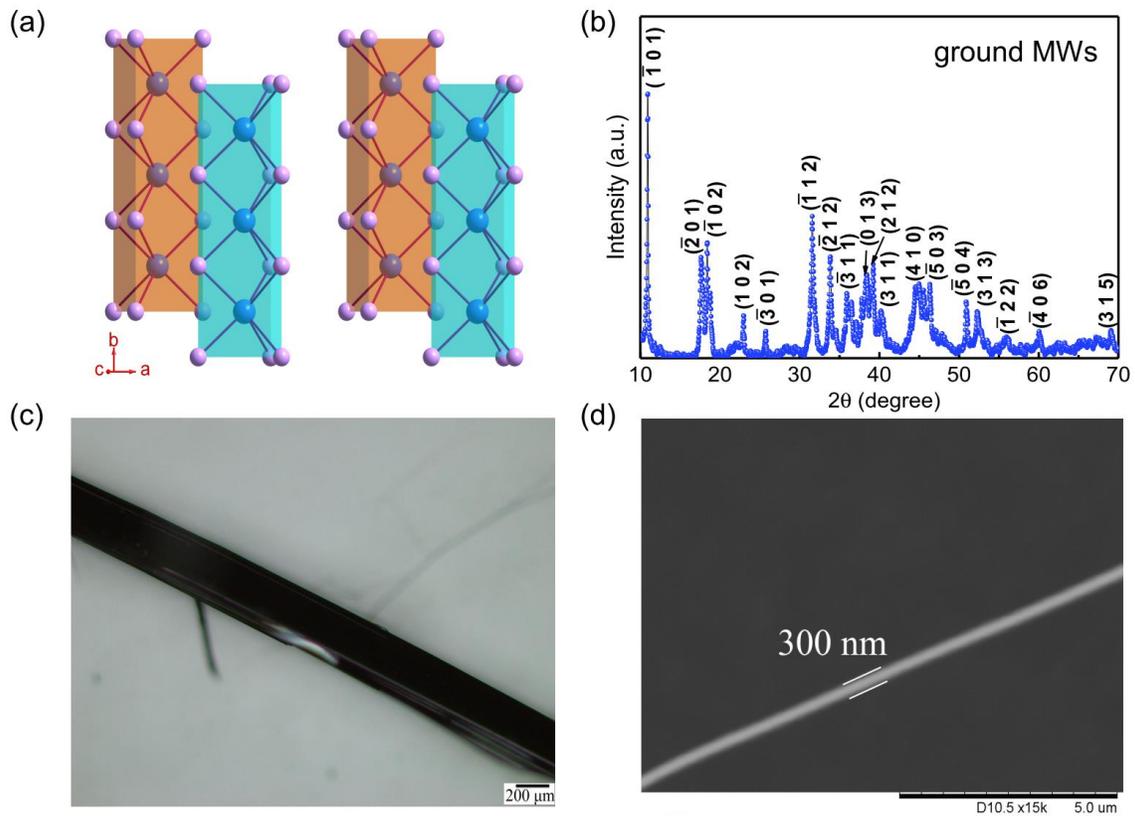

FIG. 1. (a) Schematic illustration of crystal structure of TaSe$_3$. (b) Powder XRD pattern of ground TaSe$_3$ MWs. (c) Optical image of bulk ribbon-like single crystal TaSe$_3$. (d) SEM image of a typical TaSe$_3$ MW (~300 nm thick).

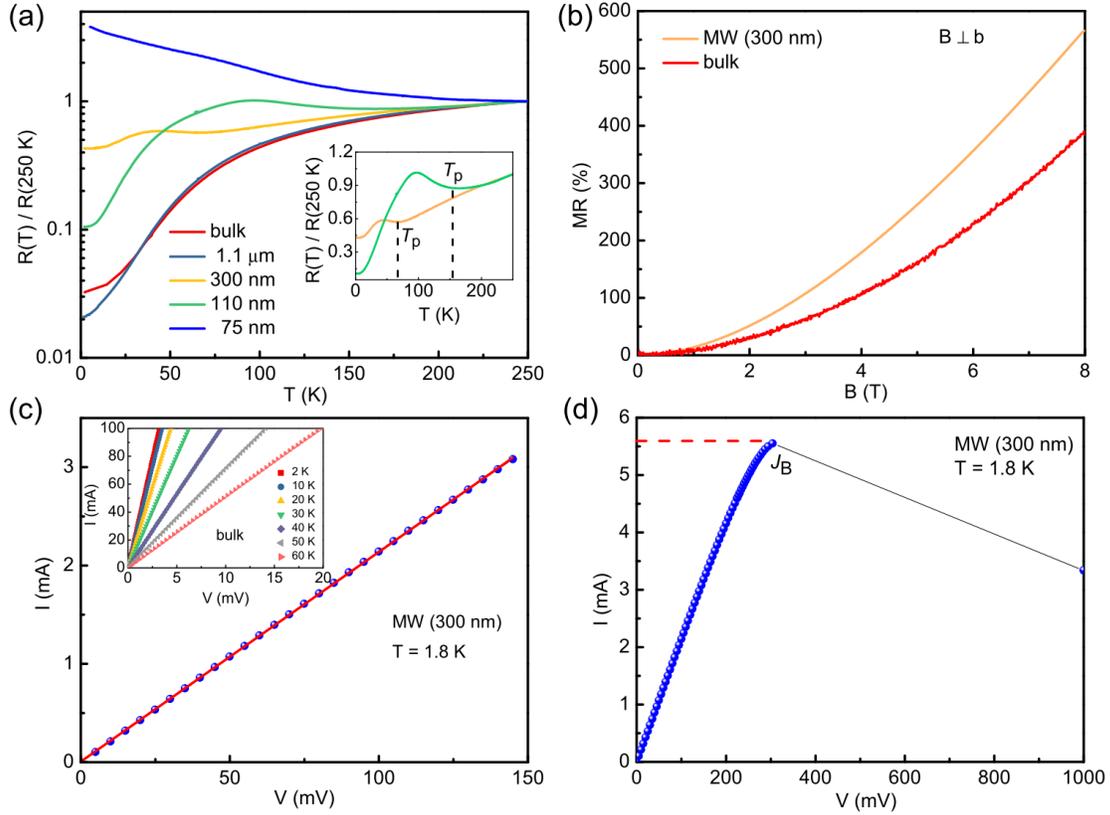

FIG. 2. (a) Temperature dependence of normalized electrical resistance for bulk and MW TaSe$_3$ samples with different diameters. Inset: *R-T* curves of the 300 nm and 110 nm MWs around the transition temperatures. (b) Magnetoresistance of bulk and MW (300 nm) samples taken at 2K in a field range of 0-8 T. (c) *I-V* curve of TaSe$_3$ MW (300 nm) taken at 1.8 K. Red line represents the fit to the Ohm's law. Inset: *I-V* curves of bulk TaSe$_3$ taken at various temperatures. (d) *I-V* curve of TaSe$_3$ MW (300 nm) taken at high electrical fields at 1.8 K.

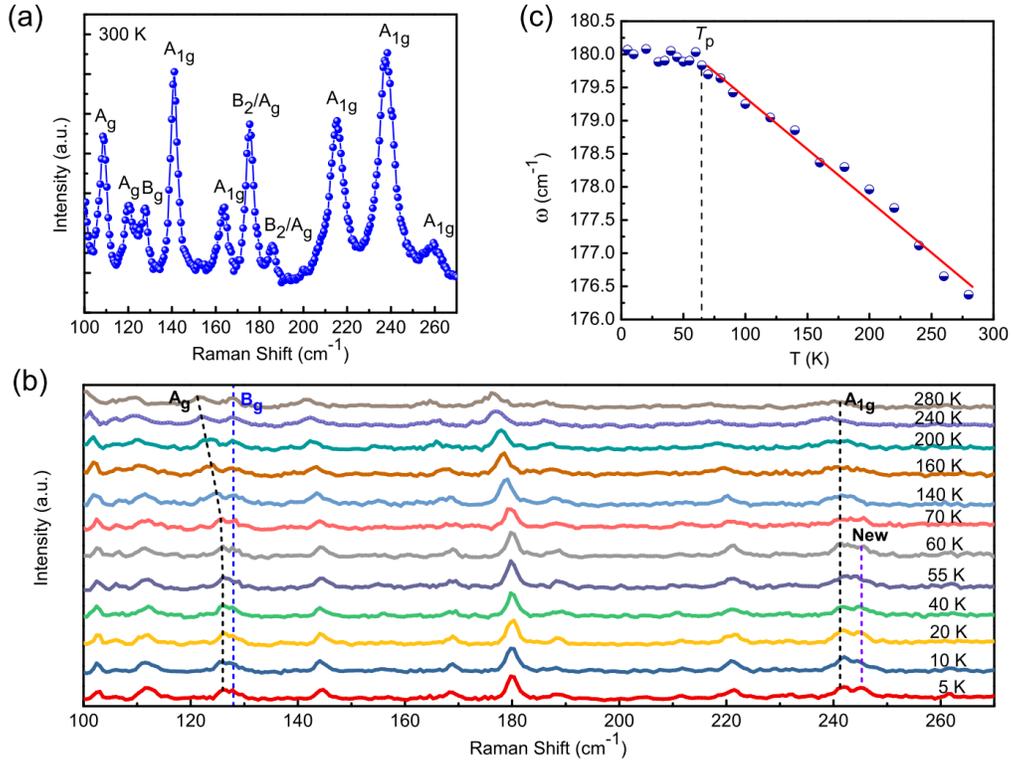

FIG. 3. (a) Raman spectrum of TaSe$_3$ MW (300 nm) taken at room temperature. (b) Raman spectra taken at various temperatures from 280 K down to 5 K. (c) Frequency of the B$_2$/A$_g$ mode as a function of temperature. Solid line represents the fit to the equation $\omega(T) = \omega_0 + \chi T$.

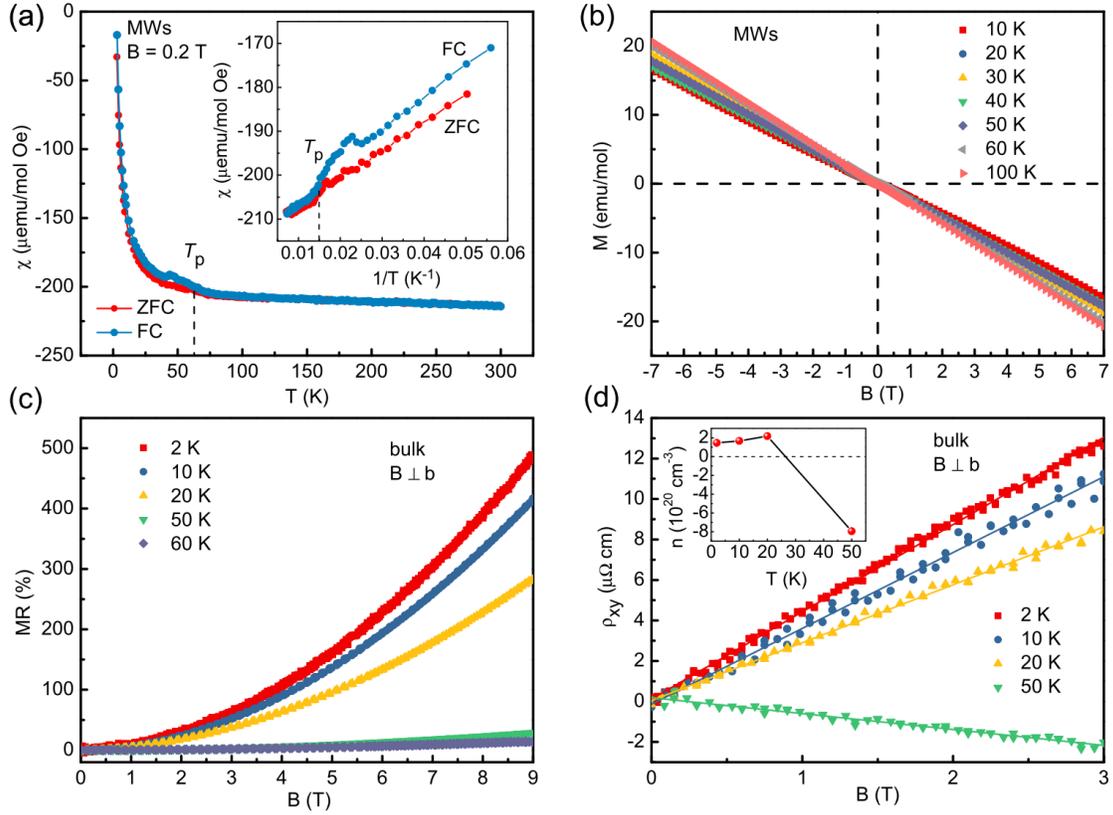

FIG. 4. (a) Temperature dependent magnetic susceptibility of TaSe$_3$ MWs taken under a field of 0.2 T. Inset: susceptibility vs inverse temperature below 100 K. (b) Magnetization of TaSe$_3$ MWs taken at various temperatures within $\pm 7$ T. (c) Magnetoresistance of bulk TaSe$_3$ taken at various temperatures in a field up to 9 T. (d) Hall resistivity measured below 50 K for bulk TaSe$_3$. Solid lines are the linear fits. Inset: carrier concentration as a function of temperature.

# Supplementary material for "Observation of charge density wave transition in TaSe$_3$ mesowires"

Yang et al.

In this supplementary file, we present images and data that are not included in the main text and do not affect understanding.

1. SEM images of TaSe$_3$ MWs and characterizations
2. SEM image of device
3. Raman spectra

**1. SEM images of TaSe$_3$ MWs and characterizations**

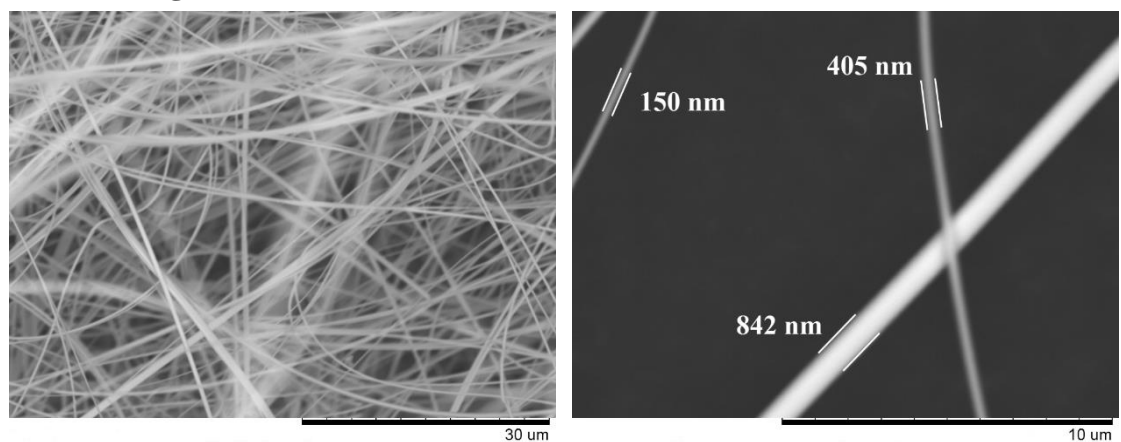

FIG. S1. SEM images of the as-grown TaSe$_3$ MWs with various diameters.

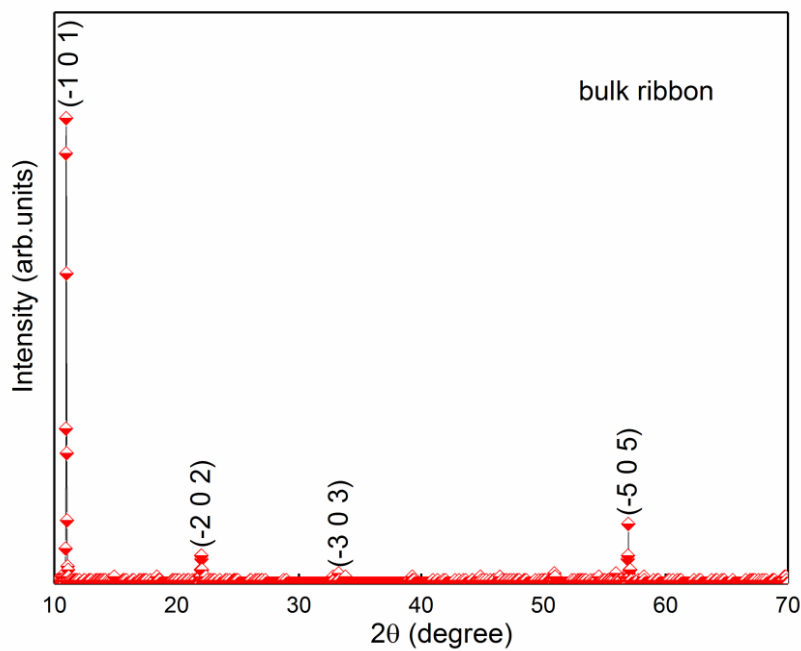

FIG. S2. Single crystal XRD pattern of bulk TaSe$_3$.

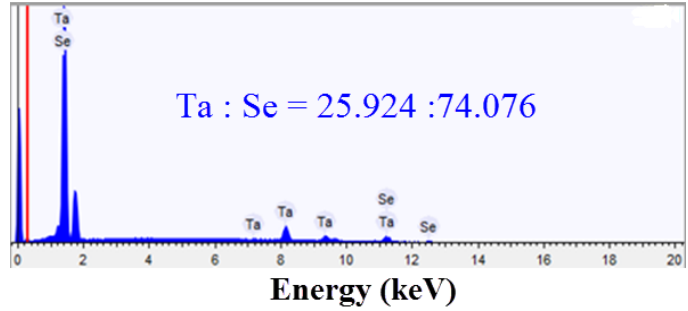

FIG. S3. EDS result of TaSe$_3$ MWs.

**2. SEM image of device**

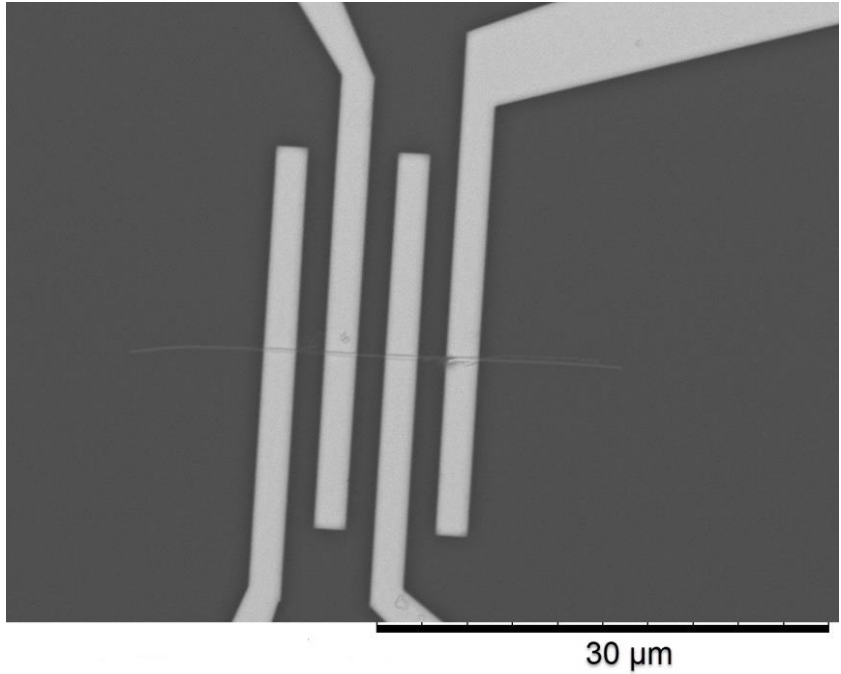

FIG. S4. SEM image of a typical device with four probes used for electrical transport measurements.

The contact and contact geometry are important to the performance of device. As was done in Ref. [1], we use a standard four probe technique to measure the resistivity of all samples. Figure S4 shows the SEM images of a typical device. The contacts are prepared through electron beam lithography followed by Au (120 nm)/Ti (10 nm) evaporation and lift-off process. All the devices show very good contact, as evidenced by the linear Ohmic behavior in the *I-V* measurements [Fig. 2(c)]. The high quality of contacts is a guarantee of reliable electrical measurements.

**3. Raman spectra**

To check the homogeneity of TaSe$_3$ MWs, we take Raman spectra at different locations on the same sample. As shown in Fig. S5, all the spectra are highly resolved and consistent, suggesting good homogeneity. Hence, the structural inhomogeneity and defects in our samples are negligibly small.

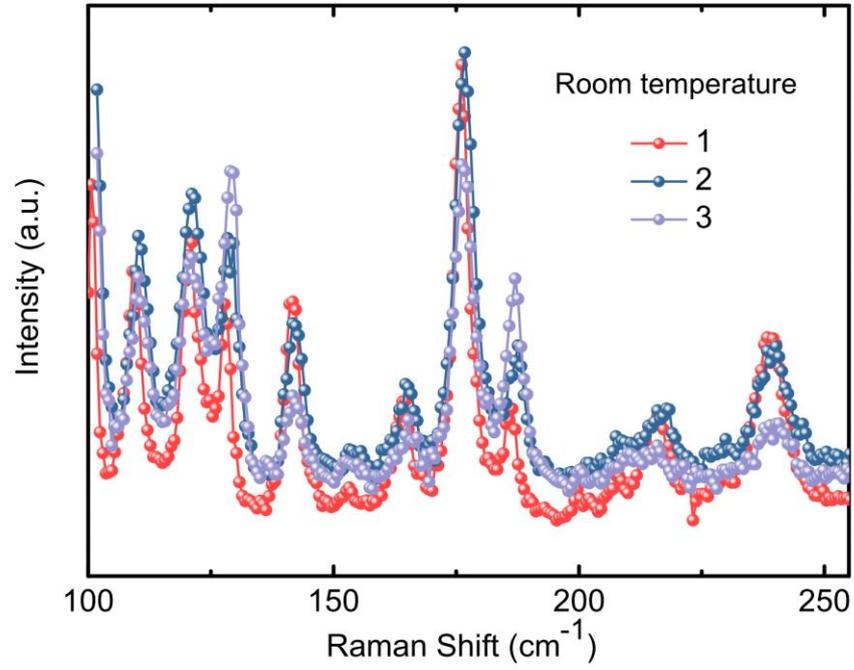

FIG. S5. Raman spectra taken at three different locations on the same sample.

[1] A. F. Isakovic, K. Cicak, and R. E. Thorne, Phys. Rev. B **77**, 115141 (2008).